\documentclass[journal,twocolumn]{IEEEtran}
\ifCLASSINFOpdf
\else
\fi
\usepackage{epsfig}
\usepackage{epstopdf}
\usepackage{mathrsfs}
\usepackage{graphicx}
\usepackage{amsfonts}
\usepackage{amsmath}
\usepackage{amssymb}
\usepackage[pdftex,bookmarksnumbered,bookmarksopen,colorlinks,citecolor=blue,linkcolor=blue]{hyperref}
\usepackage{algorithm}
\usepackage{soul}
\usepackage{cancel}
\usepackage{epstopdf}

\newcommand{\banghaiR}[1]{{\color{red} #1 }}
\newcommand{\banghaiB}[1]{{\color{blue} #1 }}
\newcommand{\banghaiY}[1]{{\color{yellow} #1 }}
\usepackage[numbers,sort&compress]{natbib}
\usepackage{amssymb, amsthm, amsmath}
\usepackage{graphicx}
\def\<{\langle}
\def\>{\rangle}

\newcommand{\comment}[1]{}

\begin{document}
%
\title{Internal Boundary between Entanglement and Separability Within a Quantum State}

\author{\IEEEauthorblockN{Bang-Hai Wang}\\
\IEEEauthorblockA{School of Computer Science and Technology, Guangdong University of Technology, Guangzhou 510006, China \\ Department of Physics, University of Oxford, Parks Road, Oxford, OX1 3PU, UK\\
Email: wangbanghai@gmail.com}
}


%


\maketitle

\begin{abstract}
Quantum states are the key mathematical objects in quantum mechanics \cite{pusey2012reality}, and entanglement lies at the heart of the nascent fields of quantum information processing and computation \cite{arute2019quantum}. However, there has not been a general, necessary and sufficient, and operational separability condition to determine whether an arbitrary quantum state is entangled or separable.


In this paper, we show that whether a quantum state is entangled or not is determined by a threshold within the quantum state. We first introduce the concept of \emph{finer} and \emph{optimal} separable states based on the properties of separable states in the role of higher-level witnesses \cite{wang2018entangled}. Then we show that any bipartite quantum state can be decomposed into a convex mixture of its optimal entangled state and its optimal separable state. Furthermore, we show that whether an arbitrary quantum state is entangled or separable, as well as positive partial transposition (PPT) or not, is determined by the robustness of its optimal entangled state to its optimal separable state with reference to a crucial threshold. Moreover, for an arbitrary quantum state, we provide operational algorithms to obtain its optimal entangled state, its optimal separable state, its best separable approximation (BSA) decomposition, and its best PPT approximation decomposition while it was an open question that how to calculate the
BSA in high-dimension systems.

\end{abstract}


%

\section{Introduction}
Quantum entanglement almost fantastically accompanied the emergence of quantum mechanics \cite{horodecki2009quantum,guhne2009entanglement}. Quantum entanglement was first described by Einstein, Podolsky, and Rosen \cite{einstein1935can}, and Schr\"{o}dinger \cite{schrodinger1935gegenwartige} as apparent \emph{paradoxes} and counter-intuitive consequences of quantum mechanics. Later, in 1964 Bell \cite{bell1964einstein} introduced the so-called Bell inequality to experimentally confirm it. In 1989 Werner mathematically formulated the definition of separability, a notion that was to be the direct opposite of entanglement \cite{werner1989quantum}. A quantum state in a composite system is called separable if it can be mathematically written as a convex combination of product states, and entangled otherwise. This accurate
definition reveals an (external) boundary between entangled
states and separable states. Here we give an
internal boundary between entanglement and separability
within an arbitrary quantum state. We show that an arbitrary bipartite quantum state can be decomposed into a linear combination of a purely entangled structure and a purely separable structure, which we call them the optimal entangled state \cite{wang2018entangled} of the quantum state and the optimal separable state of the quantum state, respectively.

Lying at the heart of quantum information theory, quantum entanglement plays a crucial role in quantum information processing
such as quantum cryptography \cite{ekert1991quantum}, quantum dense coding \cite{bennett1992communication}, quantum teleportation \cite{bennett1993teleporting}, quantum computation \cite{shor1994algorithms}, etc. Detecting and quantifying quantum entanglement, therefore, become
one of the central problems in the emerging quantum areas. Despite a number of necessary \cite{peres1996separability,horodecki1996quantum,horodecki1997separability,horodecki1999reduction,nielsen2001separable,rudolph2003some,rudolph2004computable,chen2003matrix,guhne2007covariance} or sufficient \cite{braunstein1999separability,pittenger2000separability} separability criteria, as well as the necessary and sufficient separability criterion for low dimension states \cite{HORODECKI1996separability1}, there has not been a general, necessary and sufficient operational separability condition for an arbitrary state yet. Here we classify all quantum states into disjoint families, each with a single
optimal entangled state and a single optimal separable state at its core. We show that whether an arbitrary quantum state is entangled or separable, as well as having positive partial transposition (PPT) or not, is determined by the ratio of its optimal entangled state to its optimal separable state. In other words, whether an arbitrary quantum state is entangled or separable is determined by referring to a crucial member of its family.

\textbf{Structure of the Paper.} In Section II, we provide preliminaries of the internal structure and decomposition of a quantum state and a hierarchy of witnesses.
In Section III, we give the investigation on the boundary between optimal entanglement and optimal separability within a quantum state.
In Section IV, we show what determines whether an arbitrary quantum state is entangled or separable. 
In Section V, we show the boundary between PPT entanglement and non-PPT entanglement within a quantum state.
In Section VI, we give an investigation on internal structure of entanglement and separability for the multipartite scenario.
In Section VII, we give a general algorithm of its optimal entangled state and its optimal separable state and a general algorithm of its BSA decomposition for an arbitrary quantum state. Then we fully illustrate our results using the Horodecki states \cite{horodecki1999bound}.
Finally, we offer conclusions and further study in Section VIII.

\section{Preliminaries}
For the convenience of the reader, we briefly recall the research on the internal structure and decomposition of a quantum state and a hierarchy of witnesses.
For more details, we refer to \cite{horodecki2009quantum,guhne2009entanglement,wang2018entangled}.

\subsection{The internal structure and decomposition of a quantum state}
The internal structure and decomposition of a quantum state was originally introduced in \cite{elitzur1992quantum} in the context of nonlocality, and it was later
independently rediscovered by Lewenstein and Sanpera \cite{lewenstein1998separability} in the context of entanglement and separability \cite{ducuara2020operational}. Lewenstein and Sanpera 
showed that any quantum state $\rho$ 
can always be written in a form as $\rho=\lambda\rho^{BSA}+(1-\lambda)\rho^{E}$, where $\rho^{E}$ is an entangled state, $\rho^{BSA}$ is a separable state and the weight $\lambda$ of the separable part is maximal. Later, the form was proven to be unique \cite{karnas2001separable,wang2018entangled}. The separable state $\rho^{BSA}$ is called the BSA of $\rho$, and the convex decomposition is called BSA decomposition (also called Lewenstein-Sanpera decomposition (LSD)).
On the one hand, the BSA decomposition allowed for the derivation of many very strong results \cite{karnas2001separable,sanpera1998local,kraus2000separability,horodecki2000operational,lewenstein2000optimization}.
The uniqueness of the BSA decomposition has been developed into important separability criteria in $2\times N$ bipartite system as well as $N\times N$ \cite{kraus2000separability}. Not only a remark connection between the BSA and the concurrence $C(\rho)$ was built  \cite{wellens2001separable} but also a connection between the BSA and the max- relative entropy was found \cite{quesada2014best}. On the other hand, a plethora of works
have developed on this topic \cite{quesada2014best,englert2001remarks,jafarizadeh2004best,jafarizadeh2005derivation,akhtarshenas2004optimal,thiang2009optimal,thiang2010degree,lewenstein2016sufficient}. Any convex sum of a separable state and an entangled state of a composite quantum state was called LSD in some works \cite{englert2001remarks,jafarizadeh2004best,jafarizadeh2005derivation,akhtarshenas2004optimal,thiang2009optimal,thiang2010degree}, and it was called optimal LSD when the separable state has maximal weight in the convex sum. Optimal LSD just corresponds to the notion of BSA decomposition because of the uniqueness of the BSA decomposition. For simplicity, in the following, we shall freely use optimal LSD and BSA decomposition. Many methods for the BSA decomposition were provided 
for a number of relevant types of states \cite{wellens2001separable,quesada2014best,englert2001remarks,jafarizadeh2004best,jafarizadeh2005derivation,akhtarshenas2004optimal,thiang2009optimal,thiang2010degree}. It is worth mentioning that the problem of finding the optimal LSD in some special cases \cite{jafarizadeh2004best,jafarizadeh2005derivation,thiang2009optimal} was formulated as a semidefinite program (SDP) \cite{vandenberghe1996semidefinite}. Moreover, Thiang provided a procedure to obtain the optimal LSD of a bipartite state of any finite dimension via a sequence of semidefinite relaxation \cite{thiang2010degree}. However, the maximal $\lambda$ depends on the chosen separable set in the definition of so-called optimal LSD, and there is still no one universal practical method for finding the BSA decomposition for an arbitrary state. All in all, the BSA decomposition is known for a lot of relevant types of states but, despite a good understanding of its properties,
presently we do not have an operational method for finding it for an arbitrary state. Here we give another novel decomposition of any quantum state and an algorithm for this decomposition. Compared with other methods for the BSA decomposition, this decomposition makes it easier for us to obtain the BSA decomposition by comparing the weight of its optimal entangled state with a crucial threshold. 

\subsection{The optimal entangled state of a quantum state}
For our purpose, we first consider a finite-dimensional bipartite composite Hilbert space $\mathcal{H}=\mathcal{H}_1\otimes\mathcal{H}_2$. 
The quantum state $\sigma$ in such a system is called separable if it can be written as
\begin{equation}\label{SeparableEquation}
\sigma=\sum_k{p_k|\varphi_1^k\rangle\langle\varphi_1^k|\otimes|\varphi_2^k\rangle\langle\varphi_2^k|},
\end{equation}
where $p_k$ is a probability distribution and each $|\varphi_i^k\rangle$ is a pure
state of $\mathcal{H}_i$, for $i=1,2$. If a quantum state $\rho$ cannot be written as the form of Eq. (\ref{SeparableEquation}), it is referred to as entangled.

The method of entanglement witnesses is arguably the most powerful method for entanglement detection both experimentally and
theoretically (see, e.g., \cite{chruscinski2014entanglement,wang2014characterization,augusiak2011structural} and references therein). Since the set of all separable states is convex and compact, there must exist a hyperplane that separates an arbitrary given
entangled state from the set of all separable states by the
Hahn-Banach theorem \cite{HORODECKI1996separability1,edwards2012functional}. We call this hyperplane an entanglement
witness \cite{terhal2000bell}, as shown in Fig. 1. A hermitian observable $W$ is said to be an entanglement witness if: (i) it has non-negative
expectation value for an arbitrary separable state $\sigma$; and (ii) it is not semi-positive. For an entanglement witness $W$, we can define a set of states in which the expectation value of the entanglement witness is negative, i.e. the entangled state $\rho$ is detected by $W$. We denote this set by $D_W=\{\rho|tr(W\rho)<0\}$. We say that $W_2$ is finer than $W_1$ if $D_{W_1}\subseteq D_{W_2}$, namely, if all the entangled states detected by $W_1$ are also detected by $W_2$. It was shown that $W_2$ is finer than $W_1$ if and only if there exists a $P\ge0$ and $0\le\epsilon<1$ such that $W_1=(1-\epsilon)W_2+\epsilon P$. It is determined that $W$ is optimal if there is no other entanglement witness which is finer. We can obtain that $W$ is optimal if and only if for all $P$ and $\epsilon>0$, $W'=(1+\epsilon)W-\epsilon P$ is not an entanglement witness. For more details, we refer to \cite{lewenstein2000optimization}.

For every entangled state, there is at least one entangled state to detect it and there exists an optimal entanglement witness for each entanglement witness. It is, therefore, very significant to characterize the set of optimal entanglement witnesses since optimal entanglement witnesses are sufficient to detect all the entangled states.  Although there has been a plethora
of effort in this direction, complete characterization of optimal entanglement witnesses is far from satisfactory (see, e.g., \cite{chruscinski2014entanglement,wang2014characterization,augusiak2011structural,ha2012optimal} and references therein).

Recently, we showed a hierarchy of
witnesses where entangled states play the role of high-level witnesses \cite{wang2018entangled}. The notion of entanglement witnesses was extended, and a hierarchy of witnesses for classes of observables was introduced. This hierarchy shows the fact that entangled states play the role of witnesses for detecting entanglement witnesses and separable states play the role of witnesses for the set of non-block-positive Hermitian operators. Fig. 1 illustrates the schematic picture. This framework
assembles many seemingly different findings with simple
arguments. It indicates that, for example, the answer to when different entanglement witnesses can detect the same entangled states \cite{wu2006different} and the answer to when different entangled states can be detected by the same entanglement witness \cite{wu2007determining} can be obtained by each other. Moreover, it indicates that we can develop the approach of investigating entanglement witnesses to investigate entangled states or separable states.

\begin{figure}[htbp]
\epsfig{file=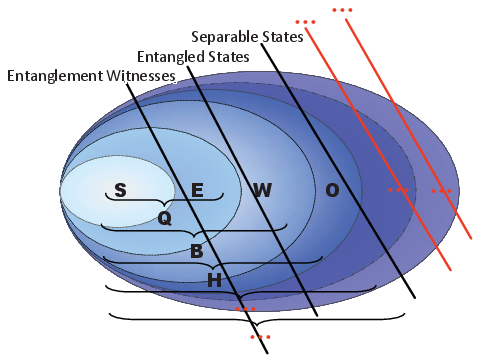,width=.95\columnwidth}
\caption{(Color online) We denote $\mathbf{S}$ the set of all separable states, $\mathbf{E}$ the set of all entangled states, $\mathbf{W}$ the set of all entanglement witnesses, $\mathbf{Q}\equiv\mathbf{S}\cup\mathbf{E}$ the set of all quantum states, $\mathbf{B}$ the block-positive operators, $\mathbf{H}$ the set of all hermitian operators, and $\mathbf{O}\equiv\mathbf{H}-\mathbf{Q}-\mathbf{W}$ the set of bounded Hermitian operators. A hierarchy structure of witnesses in which entanglement witnesses ``witness" entangled states, entangled states play the role of high-level witnesses, and separable states play the role of higher-level witnesses. }
\label{fig1}
\end{figure}

Instead of by using a numerical value \cite{brandao2005quantifying}, the entanglement of an entangled state $\rho$ was characterized by using a set of entanglement witnesses for detecting the entangled state $D_\rho=\{W|tr(W\rho)<0\}$, where $W$ is the entanglement witness of $\rho$. Given two entangled states $\rho_1$ and $\rho_2$, it was said that $\rho_2$ is finer (more entangled) than $\rho_1$ if, and only if, $D_{\rho_1}\subseteq D_{\rho_2}$, in other words, all entanglement witnesses detecting $\rho_1$ can also detect $\rho_2$. We call $\rho$ optimal if there exists no other entangled state which is finer than $\rho$. It was shown that the optimal entangled state just corresponds to the remainder (the non-separable part) \cite{karnas2001separable} of the
Lewenstein-Sanpera decomposition for a density matrix since any separable state cannot be subtracted from the optimal entangled state \cite{wang2018entangled}. Under this characterization and classification of entangled states, different entangled states belonged to different optimal-entangled-state families cannot be comparable just as we cannot generally tell which one is ``finer" from the badminton world champion (player) and the tennis world champion (player). 


\section{The boundary between optimal entanglement and optimal separability Within a Quantum State}
To characterize the partial order within entangled states, the concept of finer entangled states was introduced \cite{wang2018entangled}. And to characterize the purely entangled part of a quantum state, the concept of optimal entangled states was employed. A natural question arises. Is there a similar partial order within separable states and a purely separable part of a quantum state?
\subsection{The optimal separable state of a quantum state}
For our purpose, we have the following definitions.


\textbf{ Definitions 1.} In analogy with entanglement witnesses \cite{lewenstein2000optimization} and entangled states \cite{wang2018entangled}, we give
the following definitions about separable states. Given a separable state (higher-level witness \cite{wang2018entangled}) $\sigma$, we define $D_\sigma:=\{\Theta|tr(\Theta\sigma)<0, \Theta=\Theta^\dagger\}$; that is the set of operators ``witnessed" by $\sigma$. For our purpose, we restrict the not-block-positive Hermitian operator $\Theta$ in the ``generally-normalized" scope with $-\mathbb{I}\leq\Theta\leq \mathbb{I}$ and the operator norm $\|\Theta\|_\infty=1$ in this paper, where $\mathbb{I}$ is the identity matrix. Given two separable states, $\sigma_1$ and $\sigma_2$, we say that $\sigma_2$ is \emph{finer (more separable}) than $\sigma_1$, if $D_{\sigma_1}\subseteq D_{\sigma_2}$; that is, if all the operators ``witnessed" by $\sigma_1$, are also ``witnessed" by $\sigma_2$. We say that $\sigma$ is an \emph{optimal separable state} if there exists no other separable state which is finer than $\sigma$ (it does not exist a separable state $\sigma'$, such that $D_{\sigma}\subset D_{\sigma'}$).

However, Definitions 1 are the only trivially extended definitions of entangled states in Ref. \cite{wang2018entangled} because we have the following observations.

\textbf{Observations 1.} Consider any Hermitian operator $H$ in $\mathcal{H}=\mathbb{C}^{d_1}\otimes\mathbb{C}^{d_2}$ such that $tr(H\sigma_1)\not=tr(H\sigma_2)$ (this operator should exist if the separable $\sigma_1$ and the separable $\sigma_2$ are different density matrices). Then, let the operator $\Theta = H-tr(H\sigma_1)\frac{\mathbb{I}}{d_1d_2}$ satisfies $tr(\Theta\sigma_1)=0$, $tr(\Theta\sigma_2)=tr(H\sigma_2)-tr(H\sigma_1)\not = 0$. Hence, depending on the sign of $tr(\Theta\sigma_2)$, either $\Theta$ or $-\Theta$ does not belong to $D_{\sigma_1}$, but it belongs to $D_{\sigma_2}$. $D_{\sigma_2}\not\subset D_{\sigma_1}$. Analogously, either $H-tr(H\sigma_2)$ or $-H+tr(H\sigma_2)$ belongs to $D_{\sigma_1}$, but not to $D_{\sigma_2}$. And thus $D_{\sigma_1}\not\subset D_{\sigma_2}$.

From observations 1, we can conclude that, unfortunately, the only way that $\sigma_2$ is finer than $\sigma_1$ is that they are exactly the same state ($\sigma_2\equiv\sigma_1$) following Definitions 1. To make this partial order work, we need to employ the entangled state. \emph{In other words, the concept of the finer separable cannot independently stand on its own. From below, we can see that there can exist, exactly by referring to entangled states, 
optimal entangled states.} We have the following redefinitions.

\textbf{ Redefinitions 1.} Mathematically, we define $D_\sigma:=\{\Theta|tr(\Theta\sigma)<0, \Theta=\Theta^\dagger\}$ and $D_\Omega:=\{\Theta|tr(\Omega\Theta)\geq0, \Omega\ge0, \Theta=\Theta^\dagger\}$. Letting $D_{\sigma,\Omega}\equiv D_\sigma\cap D_\Omega$, we say that a separable state $\sigma_2$ is \emph{finer (more separable)} than a separable state $\sigma_1$, if there exists a $\Omega\ge0$ such that $D_{\sigma_1,\Omega}\subseteq D_{\sigma_2,\Omega}$; that is, if all the operators ``witnessed" by $\sigma_1$, are also ``witnessed" by $\sigma_2$. We call $\sigma$ an \emph{optimal separable state} if there exists no other separable state which is finer than $\sigma$. Note that the concept of the finer separable cannot independently exist (depending on an (optimal) entangled state $\Omega$). For simplicity, $D_{\sigma,\Omega}$ for the separable state $\sigma$ is hereinafter referred to as $D_{\sigma}$ since $D_\Omega$ is actually another constraint on $\Theta$.

Similar to the criteria for entangled states \cite{wang2018entangled} and entanglement witnesses \cite{lewenstein2000optimization}, for separable states, we have the following conclusions on the conditions if a separable state is finer than another
one, and a separable state is optimal.

\textbf{ Lemma 1.} Let the separable state $\sigma_2$ be finer than the separable state $\sigma_1$ and
\begin{equation}
\label{delta0}
\delta\equiv \inf_{\Theta_1\in D_{\sigma_1}} \left|
\frac{tr(\Theta_1\sigma_2)}{tr(\Theta_1\sigma_1)} \right|.
\end{equation}
Then we have the following:

\begin{description}
\item [(i)] If $tr(\Theta\sigma_1)=0$, then $tr(\Theta\sigma_2)\leq 0$.

\item  [(ii)] If $tr(\Theta\sigma_1)<0$, then $tr(\Theta\sigma_2) \leq
tr(\Theta\sigma_1)$.

\item [(iii)] If $tr(\Theta\sigma_1)>0$, then $\delta tr(\Theta\sigma_1)\geq tr(\Theta\sigma_2)$.

\item [(iv)] $\delta\geq 1$. In particular,
$\delta=1$ if and only if $\sigma_1=\sigma_2$.

\end{description}

{\em Proof:} Since $\sigma_2$ is finer than $\sigma_1$ we will use the
fact that for all $\Theta=\Theta^\dagger$ such that $tr(\Theta\sigma_1)<0$ then
$tr(\Theta\sigma_2)<0$.

{\em (i)} Let us assume that $tr(\Theta\sigma_2) > 0$. Then we take
any $\Theta_1\in D_{\sigma_1}$ so that for all $x\ge 0$, $\tilde \Theta(x)\equiv \Theta_1+x\Theta \in D_{\sigma_1}$ since $tr(\Theta(x)\sigma_1)=tr(\Theta_1\sigma_1)+x\cdot tr(\Theta\sigma_1)=tr(\Theta_1\sigma_1)+x\cdot0<0$ (can also see the similar proofs in Refs. \cite{wang2018entangled,lewenstein2000optimization}). But for
sufficiently large $x$ we have that
$tr(\tilde \Theta(x)\sigma_2)$
is positive, which cannot be since then $\tilde \Theta(x)
\notin D_{\sigma_2}$.

{\em (ii)} We define $\tilde \Theta=\Theta+|tr(\Theta\sigma_1)|\mathbb{I}$, where $\mathbb{I}$ is the identity
matrix. We have that ${\rm tr} (\tilde{\Theta}\sigma_1)=0$. Using (i) we have
that $0\ge
tr(\Theta\sigma_2)+|tr(\Theta\sigma_1)|$.

{\em (iii)} We take $\Theta_1\in D_{\sigma_1}$ and define $\tilde \Theta=
tr(\Theta\sigma_1) \Theta_1 + |tr(\Theta_1\sigma_1)| \Theta$, so that $tr(\tilde
\Theta\sigma_1)=0$. Using (i) we have $|tr(\Theta_1\sigma_1)|tr(\Theta\sigma_2)\le
|tr(\Theta_1\sigma_2)|tr(\Theta\sigma_1)$. Dividing both sides by
$|tr(\Theta_1\sigma_1)|>0$ and $tr(\Theta\sigma_1)>0$ we obtain
\begin{equation}
\frac{tr(\Theta\sigma_2)}{tr(\Theta\sigma_1)} \le  \left|
\frac{tr(\Theta_1\sigma_2)}{tr(\Theta_1\sigma_1)} \right|.
\end{equation}
Taking the infimum with respect to $\Theta_1\in D_{\sigma_1}$ on the
right-hand side of this equation we obtain the desired result.

{\em (iv)} By {\em (ii)}, it immediately follows that $\delta\geq1$. The ``only if" part is trivial. We prove that if $\delta=1$ then $\sigma_1=\sigma_2$.

For any positive operator $\Theta$, we have ${\rm tr}(\Theta\sigma_1)\geq0$.

Case (1): If ${\rm tr}(\Theta\sigma_1)=0$ then, by (i), ${\rm tr}(\Theta\sigma_2)=0$.

Case (2): If ${\rm tr}(\Theta\sigma_1)>0$, then by (iii)
\begin{equation}\label{Func1}
{\rm tr}(\Theta\sigma_2)\leq{\rm tr}(\Theta\sigma_1).
\end{equation}

Let $\tilde{\Theta}=-\Theta$. Then ${\rm tr}(\tilde{\Theta}\sigma_1)<0$; by Lemma 1 (ii), we have
\begin{equation}
{\rm tr}(\tilde{\Theta}\sigma_2)\leq{\rm tr}(\tilde{\Theta}\sigma_1).
\end{equation}
Hence
\begin{equation}\label{Func2}
{\rm tr}(\Theta\sigma_2)\geq{\rm tr}(\Theta\sigma_1).
\end{equation}
By Eq. (\ref{Func1}) and Eq. (\ref{Func2}), we have ${\rm tr}(\Theta\sigma_2)={\rm tr}(\Theta\sigma_1)$. According to case (1) and (2), we have, for any positive operator $\Theta$
\begin{equation}
{\rm tr}(\Theta\sigma_1)={\rm tr}(\Theta\sigma_2).
\end{equation}
Hence $\sigma_1=\sigma_2$.
$\Box$

\textbf{ Corollary 1.} $D_{\sigma_1}=D_{\sigma_2}$ if and only if $\sigma_1=\sigma_2$.

{\it Proof:} We prove the only if part. The if part is trivial. We define $\delta$ as in Eq. (\ref{delta0}) and define

\begin{equation}
\tilde{\delta}\equiv \inf_{\Theta_2\in D_{\sigma_2}} \left|
\frac{tr(\Theta_2\sigma_1)}{tr(\Theta_2\sigma_2)} \right|.
\end{equation}

By Lemma 1 (iv), we have that $\tilde{\delta}\geq1$ since $\sigma_1$ is finer than $\sigma_2$.

Equivalently, since $\sigma_2$ is finer than $\sigma_1$, we have

\begin{equation}
1\geq \sup_{\Theta_1\in D_{\sigma_1}} \left|
\frac{tr(\Theta_1\sigma_2)}{tr(\Theta_1\sigma_1)} \right|\geq\delta\geq1.
\end{equation}

Therefore, we have $\sigma_1=\sigma_2$ since $\delta=1$ according to Lemma 1 (iv).
$\Box$

\textbf{ Lemma 2.} A separable state $\sigma_2$ is finer (more separable) than $\sigma_1$ if and only if there exists
$0\leq\epsilon<1$ such that $\sigma_1=(1-\epsilon)\sigma_2+\epsilon \Omega$, where
$\Omega\geq0$ is not finer than $\sigma_1$ or $\Omega$ is a ``negative quasiprobabilities of separable state" (entangled state) \cite{sperling2009representation} such that $tr(\Omega\Theta)\geq0$ with $tr(\Theta\sigma_1)<0$ for all $\Theta=\Theta^\dag$.

{\it Proof:} (If) For all $\Theta\in D_{\sigma_1}$ we have that
$0>tr(\Theta\sigma_1)= (1-\epsilon)tr(\Theta\sigma_2)+\epsilon tr(\Theta \Omega)$
which implies $tr(\Theta\sigma_2)<0$ and therefore $\Theta\in D_{\sigma_2}$. (Only
if) We define $\delta$ as in Eq. (\ref{delta0}). Using Lemma 1(iv) we have
$\delta\ge 1$. First, if $\delta=1$ then according to Lemma 1(iv) we
have $\sigma_1=\sigma_2$ (i.e., $\epsilon=0$). For $\delta> 1$, we define
\begin{equation}
\label{deltaTop}
\hat{\delta}\equiv \sup_{\Theta_1\in D_{\sigma_1}} \left|
\frac{tr(\Theta_1\sigma_2)}{tr(\Theta_1\sigma_1)} \right|,
\end{equation}
$\Omega=(\hat{\delta}-1)^{-1}( \hat{\delta} \sigma_1-\sigma_2)$ and $\epsilon=1-1/\hat{\delta}>0$.
We have that $\sigma_1=(1-\epsilon)\sigma_2+\epsilon \Omega$ and $\hat{\delta}>1$. We can easily know that $\Omega$ is not finer than $\sigma_1$ or $\Omega$ is a ``negative quasiprobabilities of separable state" such that $tr(\Omega\Theta)\geq0$ with $tr(\Theta\sigma_1)<0$ for all $\Theta$.

Next, we prove that $\Omega$ is positive. For any $|\psi\rangle$, $\langle\psi|\Omega|\psi\rangle=(\hat{\delta}-1)^{-1}( \hat{\delta} \langle\psi|\sigma_1|\psi\rangle-\langle\psi|\sigma_2|\psi\rangle)$. Let $\Theta=-|\psi\rangle\langle\psi|$. $\frac{\langle\psi|\Omega|\psi\rangle}{|tr(\sigma_1\Theta)|}=(\hat{\delta}-1)^{-1}\frac{(tr(\sigma_2\Theta)-\hat{\delta} tr(\sigma_1\Theta))}{|tr(\sigma_1\Theta)|}=(\hat{\delta}-1)^{-1}(\hat{\delta}-\frac{|tr(\sigma_2\Theta)|}{|tr(\sigma_1\Theta)|})\geq0$.
$\Box$




\textbf{ Corollary 2.} A separable state $\sigma$ is optimal if and only if it does not exist a separable state $\sigma'=(1+\epsilon)\sigma-\epsilon \Omega$ being finer than $\sigma$ for any $\epsilon>0$ and $\Omega\geq0$ with $tr(\Omega\Theta)\geq0$ and $tr(\Theta\sigma)<0$ for all $\Theta=\Theta^\dag$.

{\em Proof:} (If) According to Lemma 2, there is no separable state which is
finer than $\sigma$, and therefore $\sigma$ is optimal. (Only if) If $\sigma'$
is a separable state, then according to Lemma 2, $\sigma$ is not optimal. 
$\Box$

We can easily conclude that the maximally mixed state is an optimal separable state by Corollary 2.

\textbf{ Corollary 3.} If $\{|\psi_i\rangle\}$ is an orthogonal (partially or completely) product basis (PB) \cite{divincenzo2003unextendible}, $\sigma=\sum_ip_i|\psi_i\rangle\langle\psi_i|$ ($p_i>0$) is an optimal separable state.

{\it Proof:}  Let us assume that $\sigma$ is not optimal. By Corollary 2, there exists at least one separable state $\sigma'=(1+\epsilon)\sigma-\epsilon \Omega$ is finer than $\sigma$, and $tr(\sigma'\Theta)<0$ for all $\Theta$ with $tr(\sigma\Theta)<0$.

Let $\Theta=t_i|\psi_i\rangle\langle\psi_i|+t\Omega$, where $t_i$ and $t$ are real numbers, and $\Omega=|\Phi\rangle\langle\Phi|$ denotes a maximally entangled state \cite{dur2000distillability}. We can obtain $tr(\sigma\Theta)=t_ip_i+t\langle\Phi|\sigma|\Phi\rangle$ and $tr(\sigma'\Theta)=(1+\epsilon)(t_ip_i+t\langle\Phi|\sigma|\Phi\rangle)-\epsilon(t_i|\langle\Phi|\psi_i\rangle|^2+t)$. Therefore, there must exist $\Theta s$ such that $tr(\sigma\Theta)<0$ and $tr(\sigma'\Theta)\geq0$ for $t_i<min\{-\frac{t\langle\Phi|\sigma|\Phi\rangle}{p_i},-\frac{t}{|\langle\Phi|\psi_i\rangle|^2}\}$. 
We can conclude that $\sigma'$ is not finer than $\sigma$, and $\sigma$ is optimal.
$\Box$

To compare the approach of the optimization over entangled states \cite{wang2018entangled} and the approach of the optimization over entanglement witnesses \cite{lewenstein2000optimization}, the optimization over separable states should be implemented by subtracting the block operator \footnote{Private communication with Marco Piani.}. 
An entanglement witness can be written as a pseudo-mixture of local projectors (product states) \cite{eckert2002entanglement}, and an entangled state can also be represented by negative quasiprobabilities of product states \cite{sperling2009representation}. To subtract this ``negative quasiprobabilities of separable state" (entangled state) from a non-optimal separable state and to keep the positivity of the resulting operator, one can only subtract the entangled state by Lemma 2. Exactly, the entangled state excluding any separable state, namely the optimal entangled state should be subtracted. However, it is still not practical. The weight of the optimal entangled state cannot easily be known because it is not the maximum number to keep the positivity of the resulting operator even if the subtracted optimal entangled state is known. Fortunately, we have an algorithm to obtain its optimal separable state and its optimal entangled state for an arbitrary state (see Section VII).

\subsection{The optimal-entanglement-and-optimal-separability decomposition of a quantum state}

Note that different from the optimal entangled state, the resulting operator may be a quantum state if we subtract an optimal entangled state from an optimal separable state, but then there exists no finer (more separable) relation between the original optimal separable state and the resulting state.
Generally, the orthogonal product basis is not unique for an optimal separable state in Corollary 2. However, the decomposition into the convex mixture of its optimal entangled state and its optimal separable state is unique for any bipartite quantum state.


\textbf{Theorem 1.} An arbitrary bipartite density matrix $\rho$ has a \emph{unique} general decomposition in the form as
\begin{equation}\label{GeneralDecompositionState}
 \rho=\Lambda\rho^{OE}+(1-\Lambda)\rho^{OS}; \Lambda\in [0,1],
\end{equation}
where (normalized) $\rho^{OE}$ denotes the optimal entangled state of $\rho$ and (normalized) $\rho^{OS}$ denotes the optimal separable state of $\rho$.

{\it Proof.---}
Case (i): $\rho$ is separable. By Lemma 2 and Corollary 2, $\rho=(1-\epsilon)\rho^{OS}+\epsilon\Omega$, where $\Omega\geq0$ is a ``negative separable state" (entangled state) such that $tr(\Omega\Theta)\geq0$ with $tr(\Theta\sigma_1)<0$ for all $\Theta=\Theta^\dag$. Without loss of generality, let $\Lambda$ denote the minimum weight 
such that $\Omega$ is positive. If $\rho$ itself is optimal separable, $\Lambda$ equals to 0. Let us assume that $\Omega$ is not optimal (entangled). There exists at least a product state $P$ and a nonnegative number $t>0$, such that $\Omega'=\Omega-tP\geq0$ by Ref. \cite{wang2018entangled}. By Lemma 2, (normalized) $\sigma=\frac{1}{1-\Lambda+t}((1-\Lambda)\rho^{OS}+tP)$ is finer than $\rho$. Since $\rho^{OS}$ is the optimal separable state of $\rho$, $\rho^{OS}$ is finer than $\sigma=\frac{1}{1-\Lambda+t}((1-\Lambda)\rho^{OS}+tP)$. By Corollary 2, there does not exist any $\Theta'={\Theta'}^\dag$ such that $tr(\sigma\Theta')<0$ with $tr(\rho^{OS}\Theta')\geq0$, and there must exist at least one $\Theta$ to satisfy $tr(\Theta\rho^{OS})<0$ and $tr(\Theta\sigma)\geq0$. We can obtain $tr(\Theta P)>0$, and $tr((r\Theta+(1-r)P)P)\geq0$ for any $0\leq r\leq1$. Let $\tilde{\Theta}_r=-(r\Theta+(1-r)P)$. There must exist a $r_0$ such that $tr(\tilde{\Theta}_{r_0}P)<0$ with $tr(\tilde{\Theta}_{r_0}\rho^{OS})\geq0$. It is impossible because $\rho^{OS}$ is optimal. Therefore, $\Omega\equiv\rho^{OE}$.

Case (ii): $\rho$ is entangled. By the BSA decomposition \cite{karnas2001separable,wang2018entangled}, $\rho=\lambda\rho^{BSA}+(1-\lambda)\rho^{OE}$, where $\lambda$ denotes the maximal number such that $\rho^{BSA}$ is separable. By case (i), $\rho^{BSA}=\Lambda_0(\rho^{BSA})^{OE}+(1-\Lambda_0)(\rho^{BSA})^{OS}$ since $\rho^{BSA}$ is separable. Therefore, $\rho=(1-\lambda)\rho^{OE}+\lambda\Lambda_0(\rho^{BSA})^{OE}+\lambda(1-\Lambda_0)(\rho^{BSA})^{OS}$. We can conclude that $(1-\lambda)\rho^{OE}+\lambda\Lambda_0(\rho^{BSA})^{OE}$ must be an optimal unnormalized entangled state, otherwise $\lambda\rho^{OE}$ can be ``consumed" and $\lambda$ is not the maximal number such that $\rho^{BSA}$ is separable. That is, despite the case the mixture of
two different single optimal entangled states might not be an optimal entangled state, it is not the case here. By the uniqueness of the optimal entangled state of an entangled state \cite{wang2018entangled}, $(\rho^{BSA})^{OE}\equiv\rho^{OE}$. By Case (i), $(\rho^{BSA})^{OS}\equiv\rho^{OS}$. Therefore, $\rho=\Lambda\rho^{OE}+(1-\Lambda)\rho^{OS}$, where $\Lambda=1-\lambda(1-\Lambda_0)$.

By Case (i) and Case (ii), we draw our conclusion. $\Box$

\begin{figure}[htbp]
\epsfig{file=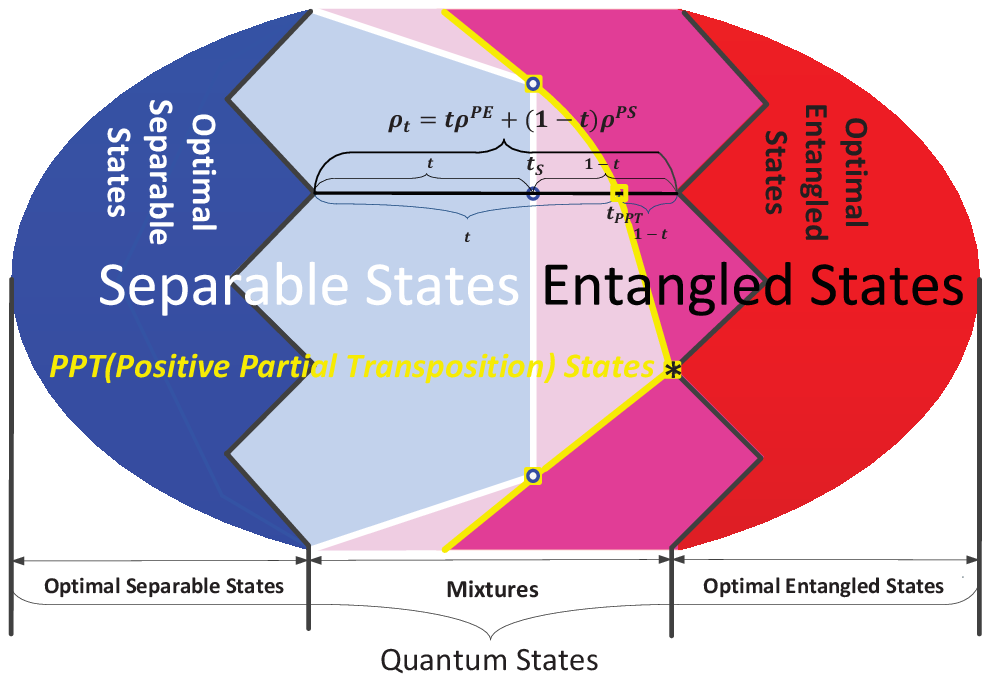,width=.95\columnwidth}
\caption{(Color online) The blue area on the far left denotes the optimal separable state, and the red area on the far right denotes the optimal entangled state. Bipartite quantum states can be classified into optimal entangled states, optimal separable states, and their convex mixtures. The boundary between separable states and entangled states is marked as a white line and the boundary between PPT states and non-PPT states is marked as a yellow line. The threshold from separability to entanglement $T_S$ and the threshold from PPT to non-PPT $T_{PPT}$ are marked as ``$\banghaiB{\circ}$" and ``$\banghaiY{\Box}$", respectively. The overlap of ``$\banghaiB{\circ}$" and ``$\banghaiY{\Box}$" denotes the overlap of $T_S$ and $T_{PPT}$, and also denotes the overlap of the boundary between separable states and entangled states and the boundary between PPT states and non-PPT states. The overlap of ``$*$" and ``$\banghaiY{\Box}$" depicts that there exist bound optimal entangled states. }
\label{fig2}
\end{figure}

\textbf{Remark 1.} This result means that bipartite quantum states can be classified into optimal entangled states, optimal separable states, and their convex mixtures. The set of bipartite quantum states is decomposed into 
families. Each family contains a single optimal entangled state and a single optimal separable state, and the other members of the family are obtained by mixing this optimal entangled state with this optimal separable state, as shown in Fig. \ref{fig2}. Note that if a single optimal entangled state is mixed with a single optimal separable state, usually the optimal entangled state (the optimal separable state) of the resulting state is not the original optimal entangled state (the optimal separable state). Consider mixing an optimal entangled state $|\psi^+\rangle=\frac{1}{\sqrt2}(|00\rangle+|11\rangle)$ with an optimal separable state $|\phi\rangle=\frac{1}{\sqrt{2}}(|10\rangle+|11\rangle)$, $\rho_m=\frac{1}{2}|\psi^+\rangle\langle\psi^+|+\frac{1}{2}|\phi\rangle\langle\phi|$. The optimal entangled state of $\rho_m$ reads $\rho_m^{OE}=|\varphi\rangle\langle\varphi|$ with the weight $\frac{3}{4}$ and the optimal separable state of $\rho_m$ reads $\rho_m^{OS}=|\varphi'\rangle\langle\varphi'|$ with the weight $\frac{1}{4}$, where $|\varphi\rangle=\frac{\sqrt6}{6}(|00\rangle+|10\rangle+2|11\rangle)$ and $|\varphi'\rangle=\frac{\sqrt2}{2}(-|00\rangle+|10\rangle)$.

\section{What determines whether an arbitrary quantum state is entangled or separable? }
Given a quantum state $\rho$, consider its family---all convex mixtures of its optimal entangled state $\rho^{OE}$ and its optimal separable state $\rho^{OS}$,
\begin{equation}\label{AnyDecomposition}
 \rho_{t}=t\rho^{OE}+(1-t)\rho^{OS}
\end{equation}
with the weight $t$ varying from 0 to 1.

\textbf{Lemma 3.} An arbitrary entangled state $\rho$ has the BSA decomposition
\begin{equation}\label{Lewenstein-Sanpera-decomposition}
\rho=\Lambda_r\rho^{OE}+(1-\Lambda_r)\rho^{BSA},
\end{equation}
where $\rho^{BSA}=\frac{\Lambda-\Lambda_r}{1-\Lambda_r}\rho^{OE}+\frac{1-\Lambda}{1-\Lambda_r}\rho^{OS}$ for $\rho=\Lambda\rho^{OE}+(1-\Lambda)\rho^{OS}$, and the remainder weight of the optimal entangled state $\Lambda_r$ is the threshold (the minimum real number, and $1-\Lambda_r$ is maximal) such that $\rho^{BSA}$ is separable.

{\it Proof.---} By Ref. \cite{wang2018entangled}, $\rho^{OE}$ (with different weight) is just the remainder of the BSA decomposition of $\rho$. It is clear that $\rho^{BSA}$ is separable because $t_S=\Lambda-\Lambda_r$ is the threshold to which $\rho_{t}=t\rho^{OE}+(1-t)\rho^{OS}$ is just separable with $t$ increasing from 0 to $t_S$. If we subtract any projector onto a product vector after we subtract $(1-\Lambda_r)\rho^{BSA}$ from $\rho$, then the resulting operator is no
longer an entangled state. By the Lemma 1 in Ref. \cite{horodecki1997separability}, any product vector in the decomposition of separability on $\rho^{BSA}$ must belong to the range of $\rho$. By the uniqueness of the BSA decomposition \cite{karnas2001separable,wang2018entangled}, we know Eq. (\ref{Lewenstein-Sanpera-decomposition}) is the BSA decomposition of $\rho$.
$\Box$

\textbf{Remark 2.} The family member will be entangled when the weight of its optimal entangled state $t$ goes beyond a threshold $t_S$, while the family member will be separable when $t$ is within the threshold, as shown in Fig. \ref{fig2}.

To illustrate this result, we sketch the proof (calculation) of the well-known threshold $p=\frac{1}{3}$ for the Werner state
\begin{equation}\label{Werner}
 \rho_p=p|\psi^+\rangle\langle\psi^+|+(1-p)\frac{\mathbb{I}}{4},
\end{equation}
 where $|\psi^+\rangle=\frac{1}{\sqrt{2}}\left(|00\rangle+|11\rangle\right)$ and $0\leq p\leq1$~\cite{werner1989quantum}. 

{\it Proof (Calculation).---} By Ref. \cite{wang2018entangled}, $|\psi^+\rangle\langle\psi^+|$ is the optimal entangled state of $\rho_p$. Suppose the Hermitian operator $\Theta$ satisfies $tr(\rho_p\Theta)<0$ and $tr(|\psi^+\rangle\langle\psi^+|\Theta)\geq0$. Therefore, $tr(\frac{\mathbb{I}}{4}\Theta)<0$, and $tr(\Theta)<0$. Without loss of generality, suppose $\Theta=t|\psi^+\rangle\langle\psi^+|+(-t-\epsilon)T$, where $T\in \{|\psi^+\rangle\langle\psi^+|\}^\bot$ and $\{|\psi^+\rangle\langle\psi^+|\}^\bot$ denotes the (orthogonal) complementary subspace of $\{|\psi^+\rangle\langle\psi^+|\}$. We can obtain $t>0$ and $\epsilon>0$ since $tr(\Theta)<0$. Thus, $tr(\rho_p\Theta)=pt+\frac{1-p}{4}(-t-\epsilon)=(\frac{3p}{4}-\frac{1}{4})t-\frac{1-p}{4}\epsilon<0$ for any $t>0$ and $\epsilon>0$. Therefore, $\frac{3p}{4}-\frac{1}{4}\leq0$, and $1\leq p\leq\frac{1}{3}$. In other words, $\frac{\mathbb{I}}{4}$ is finer (more separable) than $\rho_p$ for $0\leq p\leq\frac{1}{3}$. Similarly, we can also conclude that $\rho_q$ is \emph{finer (more separable)} than $\rho_p$ for $0\leq q\leq p\leq\frac{1}{3}$, but $\rho_q$ is \emph{finer (more entangled)} than $\rho_p$ for $\frac{1}{3}<q\leq p\leq 1$ from Ref. \cite{wang2018entangled}.
$\Box$


Let $\tau$ and $\sigma$ be the quantum states acting on a bipartite system $\mathcal{H}=\mathbb{C}^{d_1}\otimes\mathbb{C}^{d_2}$. Vidal and Tarrach \cite{vidal1999robustness} defined the robustness of $\tau$ relative to $\sigma$, $R(\tau||\sigma)$, to be the minimum nonnegative number $t$ such that the state $\rho=\frac{1}{1+t}\tau+\frac{t}{1+t}\sigma$ is separable. Interestingly, to get the BSA of a quantum state, we can employ results about the robustness of entanglement.

\textbf{Theorem 2.} An arbitrary bipartite density matrix $\rho$ has the BSA decomposition
\begin{equation}
\rho=\Lambda\rho^{OE}+(1-\Lambda)\rho^{OS},
\end{equation}
if $R(\rho^{OE}\parallel\rho^{OS})$ is infinite, otherwise
\begin{equation}
 \rho=\Lambda_r\rho^{OE}+(1-\Lambda_r)\rho^{BSA},
\end{equation}
where $\rho^{BSA}=\frac{1}{1+R(\rho^{OE}\parallel\rho^{OS})}\rho^{OE}+\frac{R(\rho^{OE}\parallel\rho^{OS})}{1+R(\rho^{OE}\parallel\rho^{OS})}\rho^{OS}$, $\Lambda_r=\frac{\Lambda(1+R(\rho^{OE}\parallel\rho^{OS}))-1}{R(\rho^{OE}\parallel\rho^{OS})}$, and $R(\rho^{OE}\parallel\rho^{OS})$ denotes the robustness of $\rho^{OE}$ relative to $\rho^{OS}$ \cite{vidal1999robustness}.

Therefore, to get the BSA of a quantum state, we can use results about the robustness of entanglement.

\textbf{Lemma 4 \cite{harrow2003robustness}.} The random robustness of a pure entangled state $|\psi\rangle$ acting on a bipartite system $\mathbb{C}^{d_1}\otimes\mathbb{C}^{d_2}$,
\begin{equation}
  R_r(|\psi\rangle)\equiv R(|\psi\rangle\langle\psi|||\frac{\mathbb{I}}{d_1d_2})=r_1r_2d_1d_2,
\end{equation} where $|\psi\rangle=\sum_jr_j|j\rangle|j\rangle$ is the Schmidt decomposition of $|\psi\rangle$ with $r_1\geq r_2\geq...\geq0$.

\textbf{Corollary 4.} For an arbitrary bipartite density matrix
$\rho=\Lambda|\psi\rangle\langle\psi|+(1-\Lambda)\frac{\mathbb{I}}{d_1d_2}$,
the BSA decomposition 
\begin{equation}
 \rho=\Lambda_r|\psi\rangle\langle\psi|+(1-\Lambda_r)\rho^{BSA},
\end{equation}
where $|\psi\rangle\langle\psi|$ is a pure entangled state, $d_1d_2$ is the dimension of the state space,
$\Lambda_r=\frac{\Lambda(1+r_1r_2d_1d_2)-1}{r_1r_2d_1d_2}$, and $\rho^{BSA}=\frac{1}{1+r_1r_2d_1d_2}|\psi\rangle\langle\psi|+\frac{r_1r_2d_1d_2}{1+r_1r_2d_1d_2}\frac{\mathbb{I}}{d_1d_2}$.

\section{What determines whether an entangled state is free or PPT?}

Similar to the BSA, we can define the best positive partial transposition approximation (BPPTA) \cite{verstraete2002geometry,bertlmann2008geometric}. In analogy to the analysis of the BSA, we can describe the properties and characterization of the BPPTA. The BPPTA can naturally serve as a quantification of entanglement. We can easily conclude that the boundary between the PPT states and the non-PPT states overlaps with the boundary of the BPPTA.
Moreover, the separable boundary and the PPT boundary, overlap in some cases. In particular, the two boundaries completely overlap in the case of low dimension (no PPT entangled state and no BPTTA).

\textbf{Theorem 3.} An arbitrary (normalized) density matrix $\rho$ with $\rho=\Lambda\rho^{OE}+(1-\Lambda)\rho^{OS}$ has a \emph{unique} decomposition in the form of
\begin{equation}
 \rho=\Lambda_R\rho^{OE}+(1-\Lambda_R)\rho^{BPPTA}; \Lambda_R\in [0,1],
\end{equation}
where $\rho^{BPPTA}=\frac{\Lambda-\Lambda_R}{1-\Lambda_R}\rho^{OE}+\frac{1-\Lambda}{1-\Lambda_R}\rho^{OS}$ is the BPPTA of $\rho$, and the remainder weight of the optimal entangled state $\Lambda_R$ is the threshold (the minimum number, and $1-\Lambda_R$ is maximal) such that $\rho^{BPPTA}$ is PPT, $\Lambda_R\leq\Lambda_r$, $t_S\le t_{PPT}=\Lambda-\Lambda_R$ for the same family, and $\Lambda_r$ is the weight in Eq. (\ref{Lewenstein-Sanpera-decomposition}), $t_{PPT}$ denotes the threshold from PPT to non-PPT.

\textbf{Remark 3.} As the weight of the optimal entangled state $t$ increases from 0 to 1 in Eq. (\ref{AnyDecomposition}), the separability of $\rho_t$ changes. A quantitative change of the weight $t$ in the mixture produces a qualitative change of the resulting state. When the weight $t$ is beyond a threshold $t_S$ (marked as ``$\banghaiB{\circ}$", as shown in Fig. \ref{fig2}), $\rho_t$ changes from a separable state to an entangled state. When the weight $t$ is beyond another threshold $t_{PPT}$ (marked as ``$\banghaiR{\Box}$", as shown in Fig. \ref{fig2}), $\rho_t$ changes from a PPT state to a non-PPT state.

A fact worth mentioning is that PPT optimal entangled states exist. Since an unextendible product basis (UPB) \cite{bennett1999unextendible,divincenzo2003unextendible} for a quantum system is an incomplete orthogonal product basis whose complementary subspace contains no product state, we can construct optimal entangled states by the complementary subspace of UPB.

\textbf{Corollary 5.} The state that corresponds to the uniform mixture on the space complementary to a UPB $\{\psi_i: i=1, \ldots, n\}$ in a Hilbert space of total dimension D
\begin{equation}
\bar{\rho}=\frac{1}{D-n}(1-\sum_{j=1}^n|\psi_j\rangle\langle\psi_j|),
\end{equation}
is a PPT optimal (bound) entangled state.

\section{The multipartite scenario}

Multipartite entanglement plays a key role in quantum computing \cite{vidal2003efficient,markham2008graph}, measurement based quantum computing \cite{raussendorf2001one}, quantum phase transition \cite{sachdev2007quantum}, and even in transport efficiency in biological systems \cite{sarovar2010quantum}. Investigating quantum many-body systems has become a fundament task \cite{wang2014universal,yu2020multipartite}. Consider a finite-dimensional composite Hilbert space $\mathcal{H}=\mathcal{H}_1\otimes\mathcal{H}_2\cdots\mathcal{H}_n$. 
The quantum state $\sigma$ in such a system is called fully separable if it can be written as
\begin{equation}\label{SeparableEquation1}
\sigma=\sum_k{p_k|\varphi_1^k\rangle\langle\varphi_1^k|\otimes\varphi_2^k\rangle\langle\varphi_2^k|\otimes\cdots\varphi_n^k\rangle\langle\varphi_n^k|},
\end{equation}
where $p_k$ is a probability distribution and each $\langle\varphi_i^k|$ is a pure
state of $\mathcal{H}_i$, for $i=1,2,\cdots,n$. If a quantum state $\rho$ cannot be written as the form of Eq. (\ref{SeparableEquation1}), it is referred to as multipartite
entangled. There are many categories of \emph{partial} separability and there is a plethora of $m$-sparable definitions in the multipartite case where the state is the product of $m$ terms. For more richer definitions of $m$-partite separabilitiy and entanglement of $m$ systems $A_1\cdots A_m$, we
refer the reader to \cite{horodecki2001separability,horodecki2009quantum,guhne2009entanglement}.

\begin{figure}[htbp]
\epsfig{file=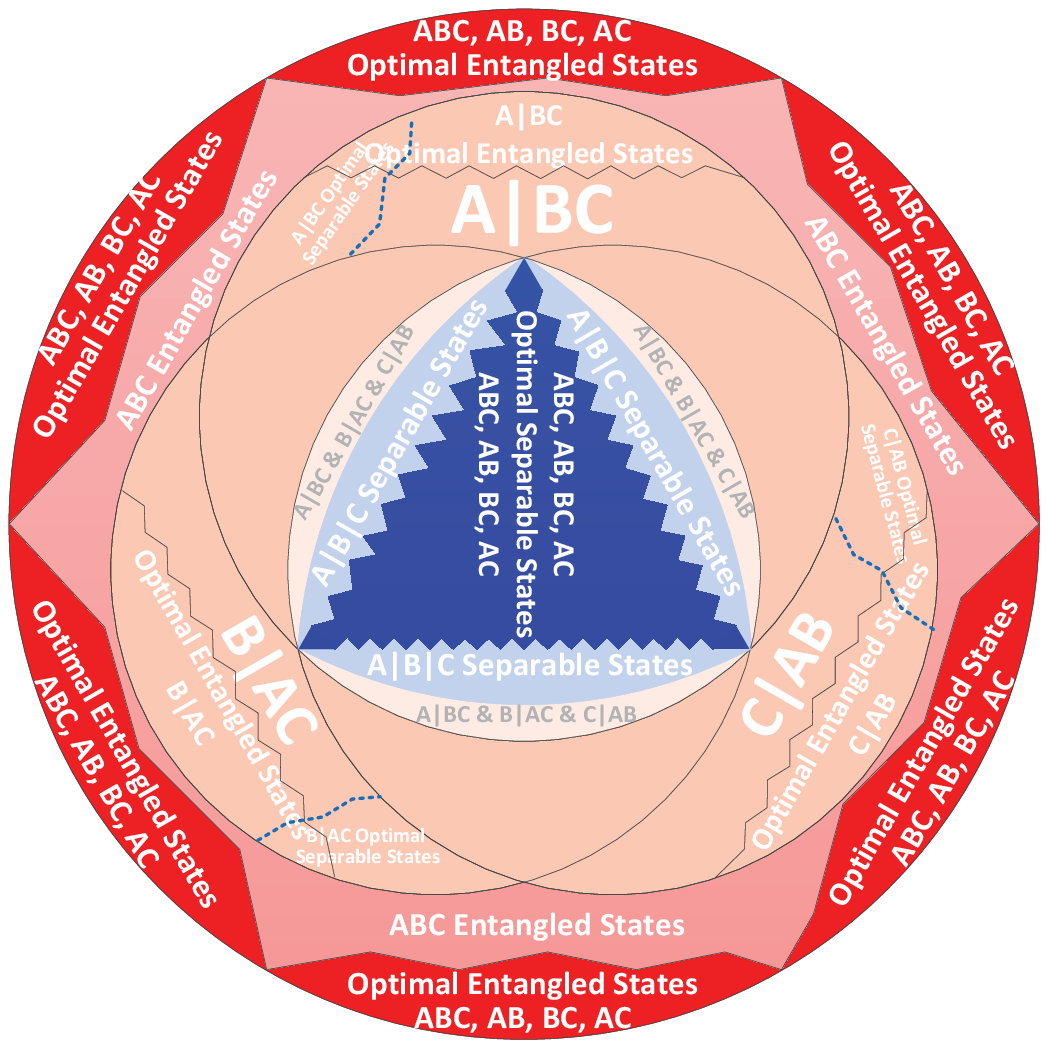,width=.95\columnwidth}
\caption{(Color online) Schematic structure of the tripartite scenario. The blue triangle depicts (full) tripartite
optimal separable states. Right next to them are full separable states. Parts marked as A-BC$\&$B-AC$\&$C-AB denote states that belong to all kinds of bipartite entangled states and that do not belong to full separable states \cite{moroder2014steering}. The outside of 3 overlapping circles denotes (genuinely) tripartite entangled states. The tripartite bipartite-separable optimal separable state and the tripartite bipartite-entangled optimal entangled state overlap. Note that the decomposition of the tripartite state isn't depicted and partial geometry properties of the tripartite state aren't shown here.}
\label{fig3}
\end{figure}

While the BSA of the multipartite diagonal symmetric state has been analytically expressed \cite{quesada2014best}, the general multipartite setting 
remains quite unexplored. Our results on the internal structure of entanglement and separability in bipartite systems can naturally be extended to the multipartite setting. However, the internal structure and decomposition of a quantum state in multipartite systems is much
richer than the bipartite systems, and the complexity of the internal structure and separability problem increases
substantially when we study multipartite systems. Since the definition and characterization of $m-$partite (full) separability in terms of positive, but not completely positive, maps and entanglement witnesses were generalized in a natural way \cite{horodecki2001separability}, we can naturally introduce the concept of $m-$partite (full) finer and optimal entangled states as well as $m-$partite (full) finer and optimal separable states. However, it is not a trivial extension of the internal structure of entanglement and separability of bipartite systems. Without loss of generality, here we consider finite-dimensional three systems $A, B, C$, i.e., $\mathcal{H}^{ABC}=\mathcal{H}_A\otimes\mathcal{H}_B\otimes\mathcal{H}_c$, there exist seven categories of optimal entangled states ($AB, BC, AC, AB|C, A|BC, B|AC, ABC$), as shown in Fig. \ref{fig3}.

\textbf{Theorem 4.} An arbitrary tripartite density matrix $\rho$ acting on $\mathcal{H}^{ABC}$ has a \emph{unique} general decomposition in the form
\begin{eqnarray}\label{TriGeneralDecompositionState}
 &\rho&=\Lambda_0\rho^{OS}_{ABC}+\Lambda_1\rho^{OS}_{AB}+\Lambda_2\rho^{OS}_{BC}+\Lambda_3\rho^{OS}_{AC}+\nonumber\\&&\Lambda_4\rho^{OS}_{AB|C}+\Lambda_5\rho^{OS}_{A|BC}+\Lambda_6\rho^{OS}_{B|AC}
 +\Lambda_7\rho^{OE}_{AB}+\nonumber\\&&\Lambda_8\rho^{OE}_{BC}+\Lambda_9\rho^{OE}_{AC}+\Lambda_{10}\rho^{OE}_{AB|C}+\Lambda_{11}\rho^{OE}_{A|BC}+\nonumber\\&&\Lambda_{12}\rho^{OE}_{B|AC}+(1-\sum_{i=0}^{12}\Lambda_i)\rho^{OE}_{ABC},
\end{eqnarray}
where $\{\Lambda_i\}_{i=0}^{12}\in [0,1]$, $\rho^{OS}_{ABC}$ denotes the (full) tripartite optimal separable state of $\rho$, $\rho^{OE}_{ABC}$ denotes the (genuinely) tripartite optimal entangled state of $\rho$, $\rho^{OS}_{AB}, \rho^{OS}_{BC}$, and $\rho^{OS}_{AC}$ denotes the bipartite optimal separable state of $\rho$, $\rho^{OE}_{AB}, \rho^{OE}_{BC}$, and $\rho^{OE}_{AC}$ denotes the bipartite optimal entangled state of $\rho$, $\rho^{OS}_{AB|C}, \rho^{OS}_{A|BC}$, and $\rho^{OS}_{C|AB}$ denote tripartite bipartite-separable optimal separable states of $\rho$, and $\rho^{OE}_{AB|C}, \rho^{OE}_{A|BC}$, and $\rho^{OE}_{C|AB}$ denote tripartite bipartite-entangled optimal entangled states of $\rho$.

We sketch the proof of Theorem 4.

{\it Proof.---} (i) According to the above analysis, for finite-dimensional $\mathcal{H}^{ABC}$, there exists a category of full (tripartite) separable states and a category of the corresponding full (tripartite) optimal separable states, which is denoted as $\rho^{OS}_{ABC}$. And there also exists a category of entangled states which cannot be written as a convex sum of projectors onto tripartite or bipartite product vectors and a category of the corresponding optimal entangled states, which is denoted as $\rho^{OE}_{ABC}$.

(ii) Referring to Ref. \cite{acin2001classification}, for finite-dimensional $\mathcal{H}^{ABC}$, there exist 6 categories ($AB, BC, AC, AB|C, A|BC, B|AC$) of bipartite separable (bipartite entangled) states, and 6 categories of the corresponding optimal bipartite separable (optimal bipartite entangled) states, which are denoted as $\rho^{OS}_{AB}, \rho^{OS}_{BC}$, $\rho^{OS}_{AC}$, $\rho^{OS}_{AB|C}, \rho^{OS}_{A|BC}$, and $\rho^{OS}_{C|AB}$ ($\rho^{OE}_{AB}, \rho^{OE}_{BC}$, $\rho^{OE}_{AC}$, $\rho^{OE}_{AB|C}, \rho^{OE}_{A|BC}$, and $\rho^{OE}_{C|AB}$), respectively.

By (i), (ii), and Theorem 1, we draw our conclusion. $\Box$

Note that $\rho^{OS}_{AB|C}$ and $\rho^{OE}_{AB|C}$ overlap because a quantum state $\rho_{AB|C}$ might be optimal separable for system AB and system C, but it might also be optimal entangled for system A and system B, 
as shown in Fig. \ref{fig3}.

\section{Algorithms and illustrations}

\subsection{A general method for an arbitrary quantum state}

Before we proceed, we need a Lemma.

\textbf{Lemma 5.} For an orthogonal (partially or completely) entangled basis \cite{divincenzo2003unextendible} $\{|\psi_i\rangle\}_{i=1}^{m}$, if the convex mixture $\rho=\sum_{i=1}^{m}k_i|\psi_i\rangle\langle\psi_i|$ is separable for any $\{k_i>0\}_{i=1}^{m}$ and $\sum_{i=1}^{m}k_i=1$, $\rho$ is (separable) optimal.




To prove Lemma 5, we need another Lemma.

 \textbf{Lemma 6 \cite{wang2018entangled,wu2007determining}.} There exists an (common) entanglement witness $W$ detected by an entangled state $\rho_1$ and an entangled state $\rho_2$ if and only if for any $k\in [0,1]$, $\rho=k \rho_1+(1-k)\rho_2$ is an entangled state.

{\it Proof.---} Without loss of generality, suppose $m$ equals 2. By Lemma 6, there exists a $k>0$ such that $\rho_{k}=k|\psi_1\rangle\langle\psi_1|+(1-k)|\psi_2\rangle\langle\psi_2|$ is separable.

Suppose $k_0$ is the minimum number such that $\rho_{k_0}=k_0|\psi_1\rangle\langle\psi_1|+(1-k_0)|\psi_2\rangle\langle\psi_2|$ is separable. Let us assume that $\rho_{k_0}$ is not optimal. So there exists at least a $\Omega>0$ such that (unnormalized) $\rho_{k_0}^\Omega=k_0|\psi_1\rangle\langle\psi_1|+(1-k_0)|\psi_2\rangle\langle\psi_2|-\epsilon\Omega$ is finer (separable) than $\rho_{k_0}$ by Corollary 2, and there exists a $\Theta=\Theta^\dagger$ such that $tr(\rho_{k_0}^\Omega\Theta)<0$ for all $tr(\rho_{k_0}\Theta)<0$.

Without loss of generality, suppose $\Theta=t_0\rho_{k_0}+t_1T_1$, where $T_1$ is contained in $\{\rho_{k_0}\}^\bot$. By $tr(\rho_{k_0}\Theta)<0$, $t_0<0$ and $t_1$ is any real number. There must exists at least one $\Theta$ with $tr(T_1\Omega)\neq0$ such that $tr(\rho_{k_0}^\Omega\Theta)\geq0$. $\rho_{k_0}^\Omega$ is not finer than $\rho_{k_0}$. Therefore, $\rho_{k_0}$ is separable optimal by Corollary 2.
$\Box$

For a given quantum state, neither its optimal entangled state nor its optimal separable state is easy to obtain. The weight of the optimal entangled state cannot easily be known because it is not the maximum number to keep the positivity of the resulting operator even if the subtracted optimal entangled state is known. Fortunately, we have a method to obtain its optimal separable state and its optimal entangled state for an arbitrary state, as shown in Algorithm \ref{Algorithm1}.

From Algorithm \ref{Algorithm1}, we can know that it is not easy for Step (iv) in Algorithm \ref{Algorithm1} because it is the procedure that, by ``consuming" entangled eigen-ensembles without any common entanglement witness, produces optimal separable states and leaves the entangled eigen-ensembles being optimal entangled states which cannot ``counteract" each other's entanglement into separability. Note that if the eigenvalues of a density matrix are degenerate, its spectral decomposition is not unique. However, the eigenspace of degenerate eigenvalues is unique. Therefore, the result of Algorithm \ref{Algorithm1} is unique.

A fact worth discussing is what the complexity of Algorithm \ref{Algorithm1} is.
Before we proceed, we need a result from Brand\~{a}o and Vianna \cite{brandao2004separable}.

 \textbf{Lemma 7 \cite{brandao2004separable}.} The determination of the optimal entanglement witness for an arbitrary state is a nondeterministic polynomial-time (NP) hard problem.

 We have the following result.

 \textbf{Theorem 5} The determination of the optimal entangled state (for an arbitrary entanglement witness) is a nondeterministic polynomial-time (NP) hard problem.

 {\it Proof.---} By Lemma 7, the determination of the optimal entanglement witness for an arbitrary state is a NP hard problem.
According to the duality of the entangled state and the entanglement witness and entangled states in the role of high-level witnesses \cite{wang2018entangled}, we conclude our result.
$\Box$

\begin{algorithm}[H]
\caption{{\bf A general method for the optimal entangled state and the optimal separable state of an arbitrary quantum state}}\label{Algorithm1}

(i) Split the eigen-ensemble \cite{hughston1993complete} of $\rho$ into two parts, the entangled eigenvectors (marked as $\{|\psi_i^E\rangle\}_{i=1}^{n_1}$ with eigenvalues $\{\lambda^{E}_i\}_{i=1}^{n_1}$, respectively) and the separable eigenvectors (marked as $\{|\psi_i^S\rangle\}_{i=1}^{n_2}$ with eigenvalues $\{\lambda^{S}_i\}_{i=1}^{n_2}$, respectively).

(ii) If there exists at least one common entanglement witness for all entangled eigenvectors, the (unnormalized) optimal entangled state of $\rho$ is $\rho^{OE}=\sum_{i=1}^{n_E}\lambda^{E}_i|\psi_i^E\rangle\langle\psi_i^E|$ and the (unnormalized) optimal separable state of $\rho$ is $\rho^{OS}=\sum_{i=1}^{n_S}\lambda^{S}_i|\psi_i^S\rangle\langle\psi_i^S|$.

(iii) Divide all the entangled eigenvectors into subsets, each containing all entangled eigenvectors without any common entanglement witness (some subsets probably contain only one entangled eigenvector)\footnotemark. 

(iv) Split each subset into the optimal entangled part and the optimal separable part.

Without loss of generality, suppose there are two (normalized) eigenvectors $|\psi^{E}_1\rangle, |\psi^{E}_2\rangle$ with eigenvalues $\lambda^{E}_1, \lambda^{E}_2$, respectively in a certain subset. Suppose $t_0$ is the minimum number such that $\rho_{t_0}=t_0|\psi^{E}_1\rangle\langle\psi^{E}_1|+(1-t_0)|\psi^{E}_2\rangle\langle\psi^{E}_2|$ is separable. According to Lemma 5, if $\frac{\lambda^{E}_1}{\lambda^{E}_2}>\frac{t_0}{1-t_0}$, $\{(\lambda^{E}_1-\frac{t_0}{1-t_0}\lambda^{E}_2)|\psi^{E}_1\rangle\langle\psi^{E}_1|\}$ is the optimal entangled part of the subset and $\{\frac{t_0}{1-t_0}\lambda^{E}_2|\psi^{E}_1\rangle\langle\psi^{E}_1|+\lambda^{E}_2|\psi^{E}_2\rangle\langle\psi^{E}_2|\}$ is the optimal separable part of the subset, else if $\frac{\lambda^{E}_1}{\lambda^{E}_2}<\frac{t_0}{1-t_0}$, $\{(\lambda^{E}_2-\frac{1-t_0}{t_0}\lambda^{E}_1)|\psi^{E}_2\rangle\langle\psi^{E}_2|\}$ is the optimal entangled part of the subset and $\{\lambda^{E}_1|\psi^{E}_1\rangle\langle\psi^{E}_1|+\frac{1-t_0}{t_0}\lambda^{E}_1|\psi^{E}_2\rangle\langle\psi^{E}_2|\}$ is the optimal separable part of the subset, else ($\frac{\lambda^{E}_1}{\lambda^{E}_2}\equiv\frac{t_0}{1-t_0}$) there is no optimal entangled part and $\{\lambda^{E}_1|\psi^{E}_1\rangle\langle\psi^{E}_1|+\lambda^{E}_2|\psi^{E}_2\rangle\langle\psi^{E}_2|\}$ is the optimal separable part.

(v) Mix all the optimal entangled parts of all subsets. The mixture just denotes the optimal entangled part of the state $\rho$. Mix all the optimal separable parts of all subsets and all separable eigenvectors in Step (i) into the optimal separable part of the state $\rho$.
\end{algorithm}
\banghaiB{\footnotetext{Note that there exists at least one common entanglement witness for the eigenvectors in \emph{different} subsets. In other words, eigenvectors within the same subset have no common entanglement witness, but eigenvectors within different subsets have at least one common entanglement witness.}}







To illustrate the algorithm, consider the Werner state. The spectral decomposition for
\begin{equation}\label{Werner}
 \rho_p=p|\psi^+\rangle\langle\psi^+|+(1-p)\frac{\mathbb{I}}{4},
\end{equation}
  reads
\begin{eqnarray}\label{WernerSpectralDecomposition}
\rho_{p}&=&\frac{1-p}{4}|\psi_{0}\rangle\langle\psi_{0}|+\frac
{1-p}{4}|\psi_{1}\rangle\langle\psi_{1}|\nonumber\\
&&+\frac{1-p}{4}|\psi_{2}\rangle\langle\psi_{2}|+\frac{1+3p}{4}|\psi_{3}\rangle
\langle\psi_{3}|,
\end{eqnarray}
where $0\leq p\leq1$, $|\psi^+\rangle=\frac{1}{\sqrt{2}}\left(|00\rangle+|11\rangle\right)$, $|\psi_{0}\rangle=|10\rangle$ and $|\psi_{1}\rangle
=|01\rangle$ are separable, while $|\psi_{2}\rangle=\frac{1}{\sqrt{2}%
}(|00\rangle-|11\rangle)$ and $|\psi_{3}\rangle=\frac{1}{\sqrt{2}%
}(|00\rangle+|11\rangle)$ are entangled. However, $|\psi_{2}\rangle$ and $|\psi_{3}\rangle$ do not have any common entanglement witness since
\begin{equation}
 \frac{1}{2}\cdot|\psi_{2}\rangle\langle\psi_{2}|+ \frac{1}{2}\cdot|\psi_{3}\rangle\langle\psi_{3}|= \frac{1}{2}(|0\rangle\langle0|\otimes|0\rangle\langle0|+|1\rangle\langle1|\otimes|1\rangle\langle1|) \nonumber
 \end{equation}
 is separable. The latter half of the Eq. (\ref{WernerSpectralDecomposition}) on the right, $\frac{1-p}{4}|\psi_{2}\rangle\langle\psi_{2}|+\frac{1+3p}{4}|\psi_{3}\rangle
\langle\psi_{3}|$, can be decomposed into $(\frac{1-p}{4}|\psi_{2}\rangle\langle\psi_{2}|+\frac{1-p}{4}|\psi_{3}\rangle
\langle\psi_{3}|=\frac{1-p}{4}(|00\rangle\langle00|+|11\rangle\langle11|))$ (the optimal separable)$+p|\psi_{3}\rangle
\langle\psi_{3}|$ (the optimal entangled). Thus, Eq. (\ref{WernerSpectralDecomposition}) can be decomposed into
\begin{eqnarray}
\rho_{p}&=&\frac{1-p}{4}|\psi_{0}\rangle\langle\psi_{0}|+\frac
{1-p}{4}|\psi_{1}\rangle\langle\psi_{1}|\nonumber\\
&&+(\frac{1-p}{4}|\psi_{2}\rangle\langle\psi_{2}|+\frac{1-p}{4}|\psi_{3}\rangle
\langle\psi_{3}|)+p|\psi_{3}\rangle
\langle\psi_{3}|\nonumber\\
&=&(1-p)\frac{\mathbb{I}}{4}+p|\psi_{3}\rangle
\langle\psi_{3}|,
\end{eqnarray}
where $\frac{\mathbb{I}}{4}$ is the optimal separable state with the weight $1-p$ and $|\psi_{3}\rangle
\langle\psi_{3}|\equiv|\psi^+\rangle\langle\psi^+|$ is the optimal entangled with the weight $p$.




\subsection{A general method for the BSA decomposition}

Now, we can get an operational method for the exact BSA decomposition, as shown in Algorithm \ref{Algorithm2}.

\begin{algorithm}[H]
\caption{{\bf A general method for the BSA decomposition}}\label{Algorithm2}



 (i) Obtain its optimal entangled state and its optimal separable state for the given quantum state by Algorithm 1.

 (ii) Calculate the threshold between the separable states
and the entangled states 
by other separability criteria (such as PPT criterion \cite{peres1996separability}, the cross-norm or realignment (CCNR) criterion \cite{rudolph2004computable,chen2003matrix}, and so on) or directly obtain the threshold by the robustness of its optimal entangled state to its optimal separable state for a given entangled state as Lemma 4.

(iii) Obtain the BSA decomposition.

\end{algorithm}

We fully illustrate our results using the Horodecki states \cite{horodecki1999bound}. It is known that
\begin{equation}\label{HorodeckiExample}
\sigma_\alpha=\frac{2}{7}|\Psi_+\rangle\langle\Psi_+|+\frac{\alpha}{7}\sigma_++\frac{5-\alpha}{7}\sigma_-,
\end{equation}
are separable for $2\leq\alpha\leq3$, bound entangled for $3<\alpha\leq4$ and free entangled for $4<\alpha\leq5$, where $|\Psi_+\rangle=\frac{1}{\sqrt3}(|00\rangle+|11\rangle+|22\rangle),\sigma_{+}={1 \over 3}(|0\rangle|1\rangle \langle 0| \langle 1|
+ |1\rangle|2\rangle \langle 1| \langle 2|+
|2\rangle|0\rangle \langle 2| \langle 0|),
\sigma_{-}={1 \over 3}(|1\rangle|0\rangle \langle 1| \langle 0|
+ |2\rangle|1\rangle \langle 2| \langle 1|+
|0\rangle|2\rangle \langle 0| \langle 2|)$.

Rewriting Eq. (\ref{HorodeckiExample}), we have
\begin{equation}
\sigma_\alpha=\frac{2}{7}P_{|\Psi_+\rangle}+\frac{5}{7}\Omega_\alpha,
\end{equation}
where $P_{|\Psi_+\rangle}=|\Psi_+\rangle\langle\Psi_+|$ and $\Omega_\alpha=\frac{\alpha}{5}\sigma_++\frac{5-\alpha}{5}\sigma_-$.
It is clear that $P_{|\Psi_+\rangle}$ is just the optimal entangled state of $\sigma_\alpha$, and $\Omega_\alpha$ is the optimal separable states of $\sigma_\alpha$.

Considering the ``big" family
\begin{equation}\label{HorodeckiState}
\sigma_\alpha^t=t P_{|\Psi_+\rangle}+(1-t)\Omega_\alpha, t\in [0,1],
\end{equation}
which include the Horodecki states, we can compute the two boundaries (thresholds) at $t_{1,2}=\frac{2\alpha^2-10\alpha-25\pm5\sqrt{4\alpha^2-20\alpha+25}}{2(\alpha^2-5\alpha-50)}$ for $0\leq\alpha\leq5$ by realigning $\sigma_\alpha^t$ according to the CCNR \cite{rudolph2004computable,chen2003matrix} 
 and the PPT boundary at $t=\frac{\alpha^2-5\alpha+5\sqrt{\alpha(5-\alpha)}}{\alpha^2-5\alpha+25}$ by positive partial transposing $\sigma_\alpha^t$ according to the PPT criterion \cite{peres1996separability} (codes available \cite{wangcode}). 
Note that from the perspective of the optimal entangled state, all states in Eq. (\ref{HorodeckiState}) belong to the family of the optimal entangled state $P_{|\Psi_+\rangle}$, but from the perspective of the optimal separable states, states in Eq. (\ref{HorodeckiState}) belong to different families with the different optimal separable states $\Omega_\alpha$ for different variables $\alpha$.

\begin{figure}[htbp]
\epsfig{file=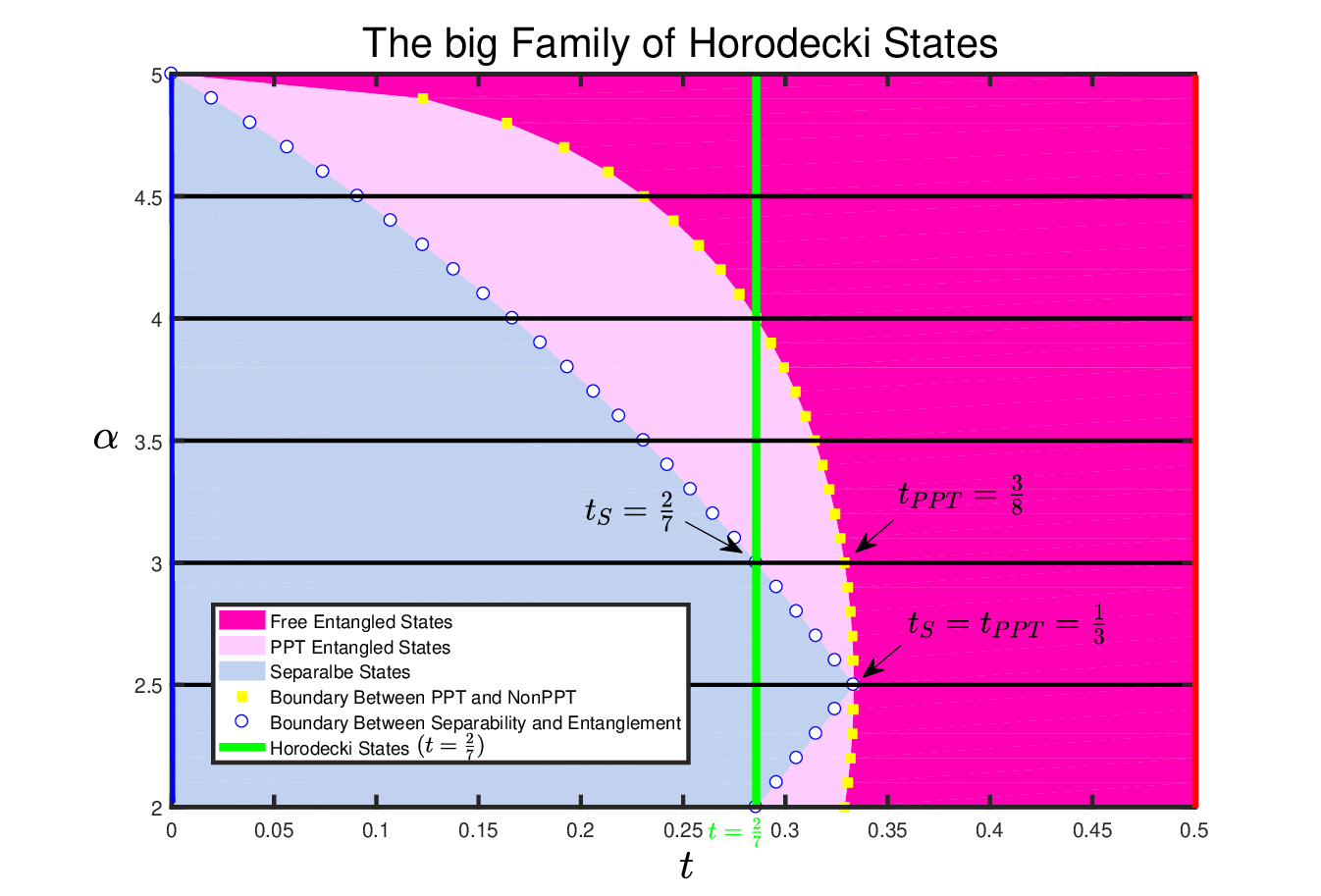,width=0.95\columnwidth}
\caption{(Color online) The separability-entanglement boundary and the PPT-NotPPT boundary in the big Family of Horodecki States. Quantum States with the weight $t=\frac{2}{7}$ of their optimal entangled state correspond to Horodecki states. The Horodecki states are separable for $2\leq\alpha\leq3$, bound entangled for $3<\alpha\leq4$ and free entangled for $4<\alpha\leq5$. Whether the other members of the big family of Horodecki states entangled or not (PPT or not) can refer to the boundary between separability and entanglement (the boundary between PPT and NonPPT). The separability-entanglement boundary and the PPT-NotPPT boundary overlap at the point $t=\frac{1}{3}$ when $\alpha=2.5$. The threshold from separability to entanglement $t_S=\frac{2}{7}$, and the threshold from PPT to NonPPT $t_{PPT}=\frac{3}{8}$ when $\alpha=3$. All the members of the big family of Horodecki states are entangled when $\alpha=5$.}
\label{fig4}
\end{figure}

Letting $\alpha=2.5$, the separable boundary overlaps with the PPT boundary, and both the BSA and the BPPTA of $\sigma_{2.5}^t$ are $\sigma_{2.5}^{\frac{1}{3}}=\frac{1}{3} P_{|\Psi_+\rangle}+\frac{2}{3}\Omega_{2.5}$ for all $t\geq\frac{1}{3}$. Letting $\alpha=3$, $t_1(t_S)=\frac{2}{7}$ and $t_2(t_{PPT})=\frac{3}{8}$, the BSA of $\sigma_3^t$ is just $\sigma_3^{\frac{2}{7}}=\frac{2}{7}P_{|\Psi_+\rangle}+\frac{5}{7}\Omega_3$ (one of the Horodecki states, as shown in Fig. \ref{fig4}) for $t\geq\frac{2}{7}$, and the BPPTA of $\sigma_3^t$ is $\sigma_3^{\frac{3}{8}}=\frac{3}{8}P_{|\Psi_+\rangle}+\frac{5}{8}\Omega_3$ for $t\geq\frac{3}{8}$.
Letting $\alpha=5$,  $\sigma_5^t=t|\Psi_+\rangle\langle\Psi_+|+(1-t)\sigma_+$
is not only the BSA decomposition but also the BPPTA decomposition of $\sigma_5^t$ because the robustness of $|\Psi_+\rangle\langle\Psi_+|$ relative to $\sigma_+$ is infinite. Fig. \ref{fig4} illustrates the schematic picture (codes available \cite{wangcode}).

\section{Conclusion}
In summary, we showed that all quantum states, entangled or separable, can be decomposed into two parts in the style of seminal Werner states, one part being its optimal entangled state, and the other part, its optimal separable state. Once a quantum state is decomposed to this form, it will be as easy as 
it is the case for the Werner state to determine whether it is entangled or not, and whether it is PPT or not. In other words, the separability or entanglement, as well as the PPT property, of a quantum state can only be determined by comparing the weight of its optimal entangled state with a crucial threshold. That the separability or entanglement of all quantum states only needs to refer to the key minority greatly promotes the separability criterion \cite{horodecki2009quantum,guhne2009entanglement}, though it didn't change the fact that the determination of any quantum state entangled or not is a nondeterministic polynomial-time (NP) hard problem \cite{gurvits2004classical,doherty2004complete}. Furthermore, we provided an operational method to obtain its optimal entangled state, its optimal separable state, its BSA decomposition, and its best PPT approximation decomposition for any finite-dimensional bipartite quantum state. How to calculate the BSA in high-dimension systems was an open question. Our results can be naturally generalized to general convex resource theories \cite{chitambar2019quantum,underpreparation}.

Here we mainly considered the case of discrete systems on the finite-dimensional Hilbert space. Our results in infinite-dimensional systems might be significantly different from the case of the discrete systems because there is no separable
neighbourhood of any mixed state in infinite-dimensional systems \cite{eisert2002quantifica}.
Our results in continuous variable systems also might be significantly different from the case of the discrete systems, because the precondition of the Hahn-Banach theorem continuous variable systems is different from the one in discrete systems \cite{wang2018entangled, bruss2002reflections}. These systems have not been discussed here. Quantum entanglement lies at the centre of quantum information and quantum computation \cite{horodecki2009quantum,guhne2009entanglement}. We hope that our findings will stimulate further investigation on quantum theory and practical applications in other fields.

We are grateful to Marco Piani, Otfried G\"{u}hne, Tristan Farrow, Zhihao Ma, Shao-Ming Fei, Xiao Yuan, and Fernando Brand\~{a}o for helpful discussions and suggestions, especially Reevu Maity, Benjamin Yadin, and Luke Hemnell for revising the original manuscript and helpful discussions and suggestions. We would also like to thank anonymous referees for their
helpful comments and suggestions to improve the original manuscript. We thank Vlatko Vedral for his kind hospitality at University of Oxford and Leong-Chuan Kwek for his kind hospitality at National University of Singapore where part of this work was carried out while Wang was an academic visitor. This work is supported by the National Natural
Science Foundation of China under Grant Nos. 62072119 and 61672007, and Guangdong Basic and Applied Basic Research Foundation under Grant No. 2020A1515011180.

\end{document}